\documentclass[prd,twocolumn,floats,floatfix,showpacs,nofootinbib,10pt]{revtex4}
\usepackage{graphicx}
\usepackage{dcolumn}
\usepackage{bm}
\usepackage{graphics}
\usepackage{slashed}
\usepackage{amssymb}
\usepackage{natbib}
\usepackage{amsmath}
\usepackage{dsfont}
\usepackage{url}
\usepackage{color}
\newcommand{\bea}{\begin{eqnarray}}
\newcommand{\ena}{\end{eqnarray}}
\newcommand{\bean}{\begin{eqnarray*}}
\newcommand{\enan}{\end{eqnarray*}}
\newcommand{\dd}{{\rm d}}

\begin{document}

\title{Disformal invariance of Maxwell's field equations}
\author{E. Goulart\footnote{egoulart@cbpf.br}, F. T. Falciano\footnote{ftovar@cbpf.br}}
\affiliation{Instituto de Cosmologia Relatividade Astrofisica ICRA -
CBPF\\ Rua Dr. Xavier Sigaud, 150, CEP 22290-180, Rio de Janeiro,
Brazil}

\date{\today}

\begin{abstract}
We show that Maxwell's electrodynamics in vacuum is invariant under active transformations of the metric. These metrics are related by disformal mappings induced by derivatives of the gauge vector $A_{\mu}$ such that the gauge symmetry is preserved. Our results generalize the well known conformal invariance of electrodynamics and characterize a new type of internal symmetry of the theory. The group structure associated with these transformations is also investigated in details.
\end{abstract}

\maketitle

\section{Introduction}

In a recent communication \cite{ertov}, we have shown that there exists a new symmetry in the relativistic wave equation for a scalar field in arbitrary dimensions. This symmetry is related to redefinitions of the metric tensor which implement a map between non-equivalent manifolds. We have encountered this result as a natural consequence of our work in analogue models of gravity, where we showed that it is possible to geometrize the dynamics of a generic nonlinear scalar field \cite{GNFT}. However, we have later realized that these metric mappings were in fact disformal transformations. Thus, in \cite{ertov} we were in fact showing that the relativistic Klein-Gordon equation is invariant under disformal transformations. 

Firstly introduced by Bekenstein in \cite{bek0, bek}, disformal transformations typically evoke the presence of an auxiliary scalar field $\psi(x)$ which appears explicitly in the transformed geometry. Thus, a disformal transformation is characterized by the relation
\begin{equation}\label{disf}
 \hat{g}_{\mu\nu}=A(\psi,\partial\psi)g_{\mu\nu}+B(\psi,\partial\psi)\partial_{\mu}\psi\partial_{\nu}\psi
\end{equation}
where $A$ and $B$ are real functions constructed with the field's invariants and $B$ is chosen to have dimensions of $M^{-4}$ so that $\psi$ has dimensions of $M$. This is the most general symmetric covariant object that can be constructed with the background metric $g_{\mu\nu}$, the scalar field $\psi$ and its first derivatives $\partial\psi$. We will call $g_{\mu\nu}$ and $\hat{g}_{\mu\nu}$ ``disformally related metrics" and the second term in the transformation as the ``disformal term'' so as to contrast to the conformal transformations that can be seen as special cases of disformal transformations with B = 0.

From their own side, conformally related geometries appear in a variety of important physical situations. From condensed matter systems to string theory, they provide a rich source of insights and are deeply ingrained in some modern field theory approaches (see, for instance, \cite{conf}). In the last decade, asymptotic and theoretical problems in quantum field theories have led to a renew of interest in conformal field theory and in the mathematical structure of the restricted conformal group SO(2,4). The SO(2,4) encompass both the Poincar\'e as well as the de~Sitter group and is a basic ingredient for the ADS/CFT correspondence \cite{Kast}.

Conformal maps and disformal transformations can be viewed as complementary concepts. While the former implements rescaling of the metric which preserve angles, the latter deforms the spacetime fabric in an anisotropic manner according to a preferred direction characterized by the gradient of the dynamical fields. Accordingly, relation (\ref{disf}) does not preserve the causal structure of the original geometry. Thus, in general, the null vectors of $g_{\mu\nu}$ are different from the null vectors of $\hat{g}_{\mu\nu}$, i.e. their cones of influence are distinct. This property has been used, for instance, to generate cosmological scenarios with varying speed of light (VSL) \cite{0}-\cite{2}. In this case, the velocity may depend not only on the functions $A$ and $B$, but also on the character of the gradient $\partial_{\alpha}\psi$ (see \cite{er} for a detailed discussion).

Since Bekentein's initial proposal, many aspects of disformal relations have been investigated. They appear in bi-metric theories \cite{san}, TeVeS models \cite{Tev}, massive gravity \cite{Mass}, DBI-Galileons \cite{dbi}, cosmic acceleration schemes \cite{disform} and others. In cosmology, for instance, they are able to reproduce many features of scalar field dark energy models: cosmological constant, quintessence, k-essence and tachyon condensates. 

The most important property of (\ref{disf}) is that it provides a natural and simple way to implement modifications of usual gravity \cite{new}. Typically, according to the disformal prescription, one replaces $g_{\mu\nu}\rightarrow\hat{g}_{\mu\nu}$ in some sector of the lagrangian and hence generates an effective coupling between the scalar field and the energy momentum tensor of the other fields describing the matter content. A fairly simple example is provided by a cosmological constant living in a disformal metric which mimics the behavior of a Chaplygin gas. The main theme of these scenarios is that there exists a duplicity of geometries. While the gravitational geometry $g_{\mu\nu}$ satisfies Einstein's equations, it is the physical geometry $\hat{g}_{\mu\nu}$ that controls the dynamics of the matter fields.

Although the role of conformal symmetries has already been extensively explored, the invariance under disformal transformations is by far less understood. It is not clear, for instance, when a given set of equations of motion are invariant under a disformal mapping. In addition, it might happen that such an invariance could help us to gain new insights in field theory as has been the case for conformal transformations.

In its seminal paper \cite{bek}, Bekenstein states that ``Maxwell's equations, the Weyl equation for spinors, gauge field equations, etc. will all be invariant under the transformation with $B$ = 0, but will not be invariant under $g_{\alpha\beta}\rightarrow \hat{g}_{\mu\nu}$ with $B\neq 0$". It is certainly true that for an arbitrary function $B$, these equations are not disformal invariant. However, in the present paper we shall show that it is possible to define a large class of disformal transformations with respect to which these theories are in fact disformal invariant.

More specifically, we shall show that Maxwell's equations in vacuum are invariant under certain disformal transformations. In dealing with electrodynamics, instead of using a scalar field such as in (\ref{disf}), the disformal transformations here introduced depend on the gauge vector $A_{\mu}$. Thus, given a metric $g_{\mu\nu}$ and the electromagnetic two-form $F_{\mu\nu}=\partial_{\mu}A_{\nu}-\partial_{\nu}A_{\mu}$ that satisfies Maxwell's equations, we will be concerned with disformal relations of the form
\begin{equation}
\hat{g}_{\mu\nu}=A(I_{1},I_{2})g_{\mu\nu}+B(I_{1},I_{2})F_{\mu}^{\phantom a\alpha}F_{\alpha\nu},
\end{equation}
where $I_{1}$ and $I_{2}$ are the electromagnetic gauge invariant scalars. In a sense, our result generalizes the usual conformal invariance of electrodynamics and constitutes a complementary internal symmetry of the theory. 

\section{Development}

In this short introductory section we shall define some relevant objects and fix our notation. Let us start with an electromagnetic field $F_{\mu\nu}$ propagating in a globally hyperbolic spacetime with metric $g_{\mu\nu}$ that has signature $(+---)$. Throughout our development, we shall consider source free field, hence, Maxwell's equations in vacuum read
\begin{equation}
g^{\mu\alpha}g^{\nu\beta}F_{ \alpha\beta ; \nu}=0 \quad , \qquad F_{[\mu\nu ;\alpha]}=0\qquad\label{max1}\\
\end{equation}
where the semicolon means covariant derivative with respect to $g_{\mu\nu}$. The second set of equations guarantee that the electromagnetic field is completely characterized by a gauge vector $A_{\mu}$, i.e. $F_{\mu\nu}=A_{[\mu;\nu]}$. Using the electromagnetic field and its dual, one can only construct two invariants, namely
\begin{eqnarray*}
I_{1}&\equiv&F^{\mu \nu}F_{\mu \nu}=2\left(H^2-E^2\right)\\
I_{2}&\equiv&\stackrel{\ast}{F^{\mu \nu}}F_{\mu \nu}=-4\vec{E}.\vec{H}
\end{eqnarray*}
where the dual bi-vector is defined as
\begin{equation}
\stackrel{\ast}{F^{\mu\nu}}=\frac{1}{2} \eta^{\mu\nu}_{\phantom a \phantom a\alpha\beta}F^{\alpha \beta},
\end{equation}
and $\eta_{\alpha\beta\mu\nu}$ is the completely antisymmetric Levi-Civita permutation tensor. The electric and magnetic fields $\vec{E}$ and $\vec{H}$ are spatial three-dimensional vector fields that are orthogonal to the observer's worldline. We also recall that the energy-momentum tensor associated with electromagnetic fields satisfying Maxwell's equations is given by
\begin{equation}\label{T}
T^{\mu\nu}=F^{\mu}_{\phantom a\alpha}F^{\alpha\nu}+\frac{I_{1}}{4}g^{\mu\nu}.
\end{equation}

One of the new feature of a disformal transformation is that the new metric may explicitly depend on the dynamical fields themselves. Note however that while the metric is a symmetric tensor, the electromagnetic two-form is antisymmetric which means that it cannot appears linearly. Thus, we need a procedure to construct a symmetric object using only the metric $g_{\mu\nu}$, the electromagnetic field $F_{\mu \nu}$ and its dual $\stackrel{\ast}{F_{\mu \nu}}$. Fortunately, due to algebraic relations between these objects, this construction is unique. Indeed, the electromagnetic field and its dual satisfy the relations
\begin{eqnarray}\label{1}
&&\stackrel{\ast}{F^{\mu \alpha}}\stackrel{\ast}{F_{\alpha \nu}}-F^{\mu \alpha}F_{\alpha \nu}=\frac{1}{2}I_{1}\delta^\mu{}_{\nu}\label{algrel1}\\\label{2}
&&\stackrel{\ast}{F^{\mu \alpha}}F_{\alpha \nu}=-\frac{1}{4}I_{2}\delta^\mu{}_{\nu}
\end{eqnarray}

Therefore, any symmetric tensor $\Delta^{\mu\nu}$ that depends only on these three fields has to be of the form
\begin{equation}\label{delta}
\Delta^{\mu\nu}=a\ g^{\mu\nu}+b\ g^{\mu\beta}g^{\nu\lambda}g^{\alpha\rho} F_{\alpha\beta}F_{\lambda\rho}\qquad,
\end{equation}
where $a$ and $b$ are two real functions that can depend on the coordinates, the electromagnetic field and its dual.

Under certain mild conditions, the quantity $\Delta^{\mu\nu}$ is invertible which allow us to use (\ref{delta}) as a disformal transformation induced by the electromagnetic tensor $F_{\mu\nu}$, i.e.
\begin{equation}\label{delta2}
g^{\mu\nu}(x)\rightarrow \Delta^{\mu\nu}(x,F_{\alpha\beta})\qquad.
\end{equation}

Note that the algebraic structure of the above disformal term is much more involved than the scalar field case. In analogy to (\ref{disf}), one could have expected to define the disformal term proportional to $\partial_{\alpha}A_\beta+\partial_{\beta}A_\alpha$. Notwithstanding, the maintenance of the gauge symmetry requires the use of the electromagnetic two-form which unavoidable leads us to (\ref{delta2}). If $b$ is zero we recover the usual conformal transformation and the causal structure of the theory is preserved. But when $b\neq0$ the disformal transformation do not preserve angles between vectors and the causal structure changes drastically. The vectors $k_{\mu}$ satisfying $\Delta^{\mu\nu}k_{\mu}k_{\nu}=0$ are, in general, not null with respect to $g^{\mu\nu}$ (see \cite{causal} for a detailed discussion) and hence the characteristic surfaces in these two situations are not the same. Only if there exist null eigenvectors of the disformal term alone the null cones of the two metrics may coincide along some specific directions. 

We shall assume that $\Delta^{\mu\nu}$ is always nonsingular, i.e. $\det(\Delta^{\mu\nu})\neq 0$. Thus, there exist a new tensor  $\Delta^{-1}_{\phantom a\mu\nu}$ such that
\begin{equation}
\Delta^{\mu\alpha}\Delta^{-1}_{\phantom a\alpha\nu}=\delta^{\mu}_{\phantom a\nu}\quad.
\end{equation}

In general, the inverse of an object of the form $g^{\mu \nu}+h^{\mu \nu}$, with arbitrary $h^{\mu\nu}$, is given as an infinite series. Notwithstanding, due to the algebraic properties encoded in the disformal term, its inverse has also a binomial form. Indeed, a direct calculation yields
\begin{equation}\label{transf2}
\Delta^{-1}_{\phantom a\alpha\nu}=Ag_{\mu\nu}+Bg^{\mu\beta}F_{\alpha\mu}F_{\beta\nu}
\end{equation}
with the coefficients $A$ and $B$ given in terms of the invariants $I_{1}$, $I_{2}$ and the previous quantities $a$, $b$
\[
A=\left(1-\frac{1}{2}pI_{1}\right)(aQ)^{-1}\quad,\qquad B=-p(aQ)^{-1}
\]
where $p\equiv b/a$ is the ``disformal ratio" and for future convenience we have defined the auxiliary quantity
\begin{equation}\label{Q}
Q\equiv1-\frac{p}{2}I_{1}-\frac{p^{2}}{16}I_{2}^{2}\quad.
\end{equation}

Having established the general form of the disformal transformation (\ref{delta}), in what follows we shall define the two up to now arbitrary functions $a$ and $b$ in such a way that the disformal transformation leaves Maxwell's equations invariant. In other words, within a large class of disformal transformations Maxwell's equations are disformal invariant.

\section{Disformal Invariance}

As it is well known, Maxwell's equations (\ref{max1}) are invariant under conformal transformations that here are described by (\ref{transf2}) with $B=0$. Thus, any electromagnetic configuration that is a solution of Maxwell's equation defined in the $g_{\mu\nu}$ manifold is also a solution of the same system of equations but in the $\Delta^{-1}_{\mu\nu}$ manifold. 

However, in general, this property will not hold if $B\neq 0$, i.e. the presence of the anisotropic stretching deforms the equations of motion in a non-trivial way. Actually, the new system of equations will depend explicitly on the choice of $A$ and $B$. Nevertheless, there is a specific choice of the function $B$ where the above mentioned property also holds, hence, by a suitable choice of $B$ any solution of Maxwell's equation in the $g_{\mu\nu}$ manifold is also a solution in the $\Delta^{-1}_{\mu\nu}$ manifold even if $B\neq 0$. 

Our first step is to calculate the action of the ``delta tensors" on the electromagnetic bi-vector. The electromagnetic two-form $F_{\mu\nu}$ is defined independently of any metric but its contravariant version does depend on which metric we are using to raise or lower the indices. To distinguish the two situation we shall use a hat over the tensor to indicate that it has been defined in the $\Delta^{-1}_{\mu\nu}$ manifold, i.e.
\begin{equation}
\hat{F}^{\mu\nu}\equiv\Delta^{\mu\alpha}\Delta^{\nu\beta}F_{\alpha\beta}\qquad.
\end{equation}
It is worth noticing that this is a highly non-linear transformation inasmuch the $\Delta^{\mu\nu}$ tensor already has a non-trivial dependency on $F_{\mu\nu}$. A straightforward calculation using (\ref{delta}) and the algebraic relations (\ref{1})-(\ref{2}) shows that $\hat{F}^{\mu\nu}$ may be written as a combination of the field and its dual as
\begin{equation}\label{poss}
\hat{F}^{\mu\nu}=\psi(I_{1},I_{2})\ F^{\mu\nu}+\chi(I_{1},I_{2})\stackrel{\ast}{F^{\mu\nu}} 
\end{equation}
with the functions $\psi$ and $\chi$ given strictly in terms of the field invariants and the pair $(a,p)$, i.e.
\begin{eqnarray*}
&&\psi=a^{2}\left[1-pI_{1}+\frac{p^{2}}{4}\left(I_{1}^{2}+\frac{I_{2}^{2}}{4}\right)\right],\\\\\\
&&\chi=a^{2}p\left(\frac{pI_{1}}{8}-\frac{1}{2}\right)I_{2}.\\
\end{eqnarray*}

The dynamical set of Maxwell's equation can be written in a more suggestive form. As long as we are only considering Riemannian manifolds we can re-write the first group of equations of (\ref{max1}) as
\begin{equation}
F^{\mu\nu}{}_{; \nu}=\frac{1}{\sqrt{-g}}\partial_\nu \left(\sqrt{-g}g^{\mu\alpha}g^{\nu\beta}F_{\alpha\beta}\right)=0 \qquad.
\end{equation}

Thus, it becomes evident that if we construct the disformal transformations such that
\begin{equation}\label{theo}
\sqrt{-g}\  g^{\mu\alpha}g^{\nu\beta}\ F_{\alpha\beta} \propto \sqrt{-\Delta}\ \Delta^{\mu\alpha}\Delta^{\nu\beta}\ F_{\alpha\beta} 
\end{equation}
then Maxwell's equations in vacuum will automatically be invariant under these transformations.

To calculate the determinant of $\Delta^{-1}_{\phantom a \mu\nu}$ we can use the Cayley-Hamilton theorem, which shows that the determinant of any mixed tensor $\textbf{T}$ can be expanded in terms of its traces as 
\begin{eqnarray*}\label{determinant}
 - 4 \, \det \textbf{T} &=& Tr (\textbf{T}^{4}) - \frac{4}{3} \,Tr (\textbf{T}) \, Tr (\textbf{T}^{3}) -\frac{1}{2} \, \left(Tr(\textbf{T}^{2})\right)^{2} +\\ 
&&+\left(Tr (\textbf{T})\right)^{2} \, Tr (\textbf{T}^{2})-\frac{1}{6} \, \left( Tr (\textbf{T}) \right)^{4}\quad.
\end{eqnarray*}

In our case, a direct calculation shows that
\begin{equation}\label{det}
\Delta\equiv \det(\Delta^{-1}_{\phantom a\mu\nu})=\det(g_{\mu\nu})\ a^{-4}\ Q^{-2}\qquad.
\end{equation}

The algebraic properties of the energy-momentum tensor have important informations about the propagation of the electromagnetic discontinuities. This local analysis can be done by studying its eigenvalue problem. It can be shown (see \cite{synge}) that the electromagnetic energy-momentum tensor has only two eigenvalues given by $\pm \kappa$ where
\begin{equation}
\kappa=\sqrt{(I_{1}^2+I_{2}^2)}\qquad.
\end{equation}
A field configuration is called algebraically general if $\kappa\neq 0$ and null if $\kappa=0$. In order to show that indeed there are disformal transformations that satisfy (\ref{theo}) we shall consider separately the algebraically general and the null cases.

\begin{itemize}
\item{General Field $\kappa\neq 0$}\\

We first note that (\ref{poss}) and (\ref{theo}) immediately imply that the term proportional to the dual must vanish so we need to impose $\chi=0$. There exist, in principle, two possibilities for $p$ that makes $\chi=0$, i.e.
\begin{equation}\label{def:pgeneralcase}
p=0\qquad \mbox{or} \qquad p=\frac{4}{I_1}
\end{equation}

Evidently, we can have $I_1=0$ but $\kappa\neq0$ so if $I_1=0$ we are force to consider $p=0$. However, $p=0$ only reproduce the conformal invariance of electrodynamics. In this manner we shall consider only the second solution which is the relevant one for our purposes. We proceed by calculating explicitly the quantity $\sqrt{-\Delta}\psi$ and imposing the condition $p=4/I_1$. Using (\ref{det}) and (\ref{Q}) we obtain that (\ref{theo}) will be satisfied if
\begin{equation}
|Q|^{-1}\left[1+(I_{2}/I_{1})^2\right]=1\quad.
\end{equation}

Incidentally, this condition is automatically satisfied. In other words, the quantity $\sqrt{-\Delta}\psi$ is constant (does not depend on the field invariants) independently on the value of the function $a$. Thus, in the case of an algebraically general field, the admissible pair is $(a,\ 4/I_1)$ with $a$ arbitrary.

\item{Null field $\kappa=0$}\\

Since the invariants appear quadratically in $\kappa$ we must have $I_{1}=I_{2}=0$. Substituting these values in $\psi$ and $\chi$, we obtain 
\begin{equation}\label{Deltak0}
\hat{F}^{\mu\nu}=a^{2}F^{\mu\nu}\quad , \quad \sqrt{-\Delta}=\sqrt{-\gamma}\ a^{-2}\quad.
\end{equation}

In the case of a null field, relation (\ref{theo}) is always satisfied independently of the particular realization of the pair $(a, p)$. Therefore, (\ref{theo}) is valid for arbitrary values of the functions $a$ and $p$.
\end{itemize}

Let us summarize our result. In terms of the gauge vector $A_{\mu}(x)$, Maxwell's equations (\ref{max1}) in vacuum may be recast in the form
\begin{equation}
\frac{1}{\sqrt{-g}}\partial_{\nu}\left(\sqrt{-g}g^{\mu\alpha}g^{\nu\beta} \partial_{[\alpha}A_{\beta]}\right)=0.
\end{equation}
If $A_{\mu}(x)$ satisfies Maxwell's equations in a spacetime endowed with metric $g_{\mu\nu}$ it also satisfies the same set of dynamical equations but in a different spacetime endowed with metric $\Delta^{-1}_{\phantom a\mu\nu}$. The metric of the latter manifold is constructed according with the proper choice of $(a,p)$ discussed above. Defining the covariant derivative``$_{||}$"such that
\begin{equation}
\Delta^{-1}_{\phantom a\mu\nu ||\alpha}=0,
\end{equation}
we immediately obtain
\begin{equation}
\Delta^{\mu\alpha}\Delta^{\nu\beta}F_{ \alpha\beta || \nu}=0, \quad\quad F_{[\mu\nu ||\alpha]}=0.
\end{equation}
Thus, in the same way that a gauge transformation $A_{\mu}\rightarrow A_{\mu}+\partial_{\mu}\Lambda$ characterizes different representations of the same physical situation, we may say that, from the formal point of view, it is impossible to distinguish between different spacetimes related by the disformal transformation given by (\ref{delta2}). In other words, all spacetimes $\Delta^{\mu\nu}$ constructed with $A_{\mu}$ and the pair of functions $(a,p)$ are compatible with the same potential configuration as a solution. This is a symmetry of Maxwell's electromagnetic theory.

\subsection{Metrical Properties of the Energy-Momentum tensor}

A remarkable property of this new symmetry is that there exist an intimate relationship between the disformal metric $\Delta^{\mu\nu}$ and the energy-momentum tensor defined in the original geometry $g_{\mu\nu}$. Let us concern ourselves to the case $\kappa\neq 0$. In this case, the disformal transformation (\ref{delta}) is given by
\begin{equation}
\Delta^{\mu\nu}=a\Big(g^{\mu\nu}+\frac{4}{I_{1}}F^{\mu}_{\phantom a\alpha}F^{\alpha\nu}\Big)\quad,
\end{equation}
with the function $a$ completely arbitrary. Making the redefinition $4a=-I_{1}\Omega^{2}$ (we choose the minus sign to keep our signature convention intact), where $\Omega$ is an arbitrary function, and using (\ref{T}) we recast the form of the admissible delta tensors as \footnote{Note that, for $\kappa\neq 0$ the energy momentum tensor is invertible, which proves the consistency of our approach.} 
\begin{equation}
\Delta^{\mu\nu}=-\Omega^{2}\ T^{\mu\nu}\quad.
\end{equation} 
In other words, the disformal metric is conformally related to the energy-momentum tensor of the original field configuration. Thus, things that look like energy and momentum in the former manifold appear as spacetime distances in the ``new" manifold. This is an interesting symmetry property of the dynamical equations that somehow generalize the concept of conformal transformations. The disformally related line element is given by
\begin{equation}
d\hat{s}^2\equiv \Delta^{-1}_{\phantom a\mu\nu}dx^{\mu}dx^{\nu}
\end{equation}

In general, the riemannian spacetimes generated by the disformal transformations  are non-flat and depend explicitly on the particular solutions of the gauge potential $A_{\mu}(x)$. 

Having defined the metrical structure of a given manifold, one can study its properties by constructing the geometrical objects and the Debever's invariants associated to them. Considering for instance the curvature tensor that has second derivatives of the metric $\Delta^{\mu\nu}$, it happens that it will also have higher derivatives of the vector potential, i.e.
\begin{equation}
\hat{R}_{\alpha\beta\mu\nu}\rightarrow \hat{R}_{\alpha\beta\mu\nu}(g,\partial g, \partial^{2}g;\  F, \partial F, \partial^{2}F)
\end{equation}

Its explicit expression is a very long and involved equation that should be analyzed for each particular solution. Besides, there seems to have no natural way to separate and classify the terms appearing in its decomposition. It is also worth noting that, in general, the metric $\Delta^{\mu\nu}$ does not have the same isometries as the original $g^{\mu\nu}$. There is no reason for these two metrics to share the same set of killing vectors.

Our analysis has focused in the invariance of the equation of motion (\ref{max1}) under metric transformations and, as it is well known, symmetries of the equation of motion do not imply symmetries in the action. However, it is straightforward to show that transformation (\ref{delta2}) is also a symmetry of the action. Indeed, the action integral
\begin{equation}
S=-\frac14\int \dd^{4}x \sqrt{-g} g^{\mu\nu}g^{\alpha\beta}F_{\mu\alpha}F_{\nu\beta}\qquad,
\end{equation}
is invariant under the map
\begin{equation}
 g_{\mu\nu}\ \rightarrow \ \Delta^{-1}_{\phantom a \mu\nu}\qquad \sqrt{-g}\ \rightarrow \ \sqrt{-\Delta}\qquad.
\end{equation}

In a Taylor expansion around a fiducial point $x_{0}$, the function $a(x)$ or $p(x)$ are characterized by an infinite number of parameters. Consequently, this symmetry of the action is also characterized by an infinite number of parameters \cite{Had}.

\subsection{A particular realization of the symmetry: static point charge}

As a simple example let us analyze the case of a motionless point-like charge. In spherical coordinates $x^{\mu}=(t, r, \theta, \phi)$, the spherically symmetric and static solution of (\ref{max1}) is of the form\\
\begin{equation}
F_{\mu\nu}=E(r)\left(\delta_{\mu}^{\phantom a 0}\delta_{\nu}^{\phantom a 1}-\delta_{\mu}^{\phantom a 1}\delta_{\nu}^{\phantom a 0}\right)
\end{equation}
where $E(r)=Q/r^{2}$. A direct calculation yields the contravariant components of the energy-momentum tensor
\begin{displaymath}
T^{\mu\nu}=\frac{E^{2}}{2}\left(
\begin{array}{cccc}
1 & 0 & 0 & 0\\
0 & -1 & 0 & 0\\
0 & 0 & r^{-2} &0\\
0 & 0 & 0& r^{-2}sin^{-2}\theta
\end{array}\right)
\end{displaymath}

For the sake of simplicity we choose $\Omega^{2}=1$. Thus,
\begin{equation}
\sqrt{det(T_{\phantom a\mu\nu}^{-1})}=\frac{16}{E^{4}}r^{2}sin^{2}\theta\qquad.
\end{equation}

A direct calculation shows that, indeed, the field satisfies the equation 
\begin{equation}
\frac{1}{\sqrt{det(T_{\phantom a\mu\nu}^{-1})}}\ \partial_{\nu}\left(\sqrt{det(T^{-1}_{\phantom a\mu\nu})}\ T^{\mu\alpha}T^{\nu\beta}\ \partial_{[\alpha}A_{\beta]}\right)=0\quad,
\end{equation}\\
which again is an explicit example of our results. The line element is given by
\begin{equation}\label{temp:ds}
\hat{ds}^{2}=\frac{2r^{4}}{Q^{2}}\left(dr^{2}-dt^{2}-r^{2}d\theta^{2}-r^{2}sin^{2}\theta \ d\phi^{2}\right)\quad.
\end{equation}

Note that the radial coordinate becomes a timelike coordinate while the time coordinate becomes spacelike. This is an intriguing property that shall be investigated in a future work. Also one can calculate the curvature tensor of (\ref{temp:ds}) and confirm that the $\Delta_{\phantom a\mu\nu}^{-1}=T_{\phantom a\mu\nu}^{-1}$ manifold is indeed curved. Notwithstanding, the conformal tensor is identically zero which means that the electric point charge generates a class of conformally flat spacetimes via the disformal mapping.

\section{Group Structure}

Another interesting property of the symmetry transformation generated by (\ref{delta2}) is that together with its set of differential manifolds they form a group for each and every solution $A_\mu(x)$ of the dynamical equations. To simplify notation we can define a symmetric tensor $\phi_{\mu \nu}$ that using the algebraic relations (\ref{1})-(\ref{2}) satisfies
\begin{eqnarray}
\phi_{\mu \nu}&\equiv&F_\mu{}^{\alpha}F_{\alpha \nu}\label{def_phi}\\
\phi_{\mu \alpha}F^\alpha{}_\nu&=&-\frac{I_2}{4}\stackrel{\ast}{F_{\mu \nu}}-\frac{I_1}{2}F_{\mu \nu}\label{phiF}\\
\phi_{\mu \alpha}\phi^\alpha{}_\nu&=&\frac{I_2^2}{16}g_{\mu \nu}-\frac{I_1}{2}\phi_{\mu \nu}\label{phiphi}
\end{eqnarray}

In addition, we will included two indices $(a,p)$ to specify the transformation. From now on we shall change a bit our notation and drop the hat used to designate the transformed objects. The indices $(a,p)$ should suffice to appropriately identify the transformation so that we write
\begin{equation}
F_{(a,p)}^{\mu\nu}\equiv\Delta_{(a,p)}^{\mu\alpha}\Delta_{(a,p)}^{\nu\beta}F_{\alpha\beta}\qquad.
\end{equation}

Once more, we shall separate our analysis in the two cases of an algebraic general or null electromagnetic field.

\subsection{General Field $\kappa \neq 0$}

For the case of an algebraic general field, eq. (\ref{def:pgeneralcase}) shows us that the disformal ratio is fix, i.e. we have only one free function. Thus, the disformal transformation reads
\begin{equation}\label{Da}
\Delta^{-1}_{(a) \mu\nu}=\frac{a^{-1}}{1+\xi^2}\left(g_{\mu\nu}+\frac{4}{I_1}\phi_{\mu \nu}\right)
\quad,
\end{equation}
where we have defined the quantity $\xi \equiv \frac{I_2}{I_1}$
The electromagnetic two-form in the $\Delta^{-1}_{(a)\mu\nu}$ geometry can be related with its counterpart in the $g_{\mu\nu}$ through
\[
F_{(a)}^{\mu \nu}=\Delta_{(a)}^{\mu\alpha}\Delta_{(a)}^{\nu\beta}F_{\alpha \beta}=a^2\left(1+\xi^2\right)g^{\mu \alpha}g^{\nu\beta}F_{\alpha \beta}\quad.
\]
Therefore it is straightforward to obtain the relations
\begin{eqnarray}
I_{1(a)}&=&a^2\left(1+\xi^2\right)I_1\label{Fa}\\
I_{2(a)}&=&a^2\left(1+\xi^2\right)I_2\\
\xi_{(a)}&=&\xi\\
\phi_{(a)\mu \nu}&\equiv&\Delta_{(a)}^{\alpha \beta}F_{\mu \alpha}F_{\nu \beta}=\frac{a I_1\xi^2}{4}g_{\mu \nu}-a\phi_{\mu \nu}\label{phia}
\end{eqnarray}

We define the transformation $\mathcal{T}$ acting on the metric $g_{\mu\nu}$ such that  
\begin{equation}\label{Transf}
\mathcal{T}_a[g_{\mu\nu}]\equiv \Delta^{-1}_{(a)\mu\nu}
\end{equation}
with $\Delta^{-1}_{(a)\mu\nu}$ defined by (\ref{Da}). According to our previous discussion the transformation symbol relates two non-equivalent manifolds. Let us apply a second transformation ${\cal T}_b$ associated with the function $b(x)$ on the metric $\Delta^{-1}_{(a)\mu\nu}$. We have
\begin{equation}
{\cal T}_b\ [{\cal T}_a\ [g_{\mu\nu}]]={\cal T}_b\ [\Delta^{-1}_{(a)\mu\nu}]\quad.
\end{equation}

Replacing all $g_{\mu\nu}$ by $\Delta^{-1}_{(a)\mu\nu}$ into (\ref{Da}) one immediately obtains
\[
{\cal T}_b\ [\Delta^{-1}_{(a)\mu\nu}]=
\frac{b^{-1}}{1+\xi^2_{(a)}}\left(\Delta^{-1}_{(a)\mu\nu}+\frac{4}{I_{1(a)}}\phi_{(a)\mu \nu}\right)\quad.
\]

A direct calculation using explicitly (\ref{Da})-(\ref{phia}) gives us
\[
{\cal T}_b\ [{\cal T}_a\ [g_{\mu\nu}]]=\frac{(a.b)^{-1}}{1+\xi^2}g_{\mu\nu}\quad,
\]
which is conformally related to the $g_{\mu \nu}$ metric. The transformation (\ref{Transf}) is homogeneous of order 1, hence any metric  conformally related to $g_{\mu \nu}$ satisfies
\[
{\cal T}_a\ [\lambda g_{\mu\nu}]=\lambda{\cal T}_a\ [g_{\mu\nu}]=\lambda\Delta^{-1}_{(a)\mu\nu} \quad.
\]

Therefore, an arbitrary number of successive transformations will generate only two types of metric, namely
\begin{eqnarray}
M_{(a)\, \mu\nu}&\equiv&a g_{\mu \nu}\quad ,\\
N_{(a)\, \mu\nu}&\equiv&a\left(g_{\mu \nu}+\frac{4}{I_1}\phi_{\mu \nu}\right)\quad.
\end{eqnarray}

The collection of all metrics $M_{(a)\, \mu\nu}$ and $N_{(a)\, \mu\nu}$ together with transformation (\ref{Transf}) can be viewed as a representation of a group $\mathcal{G}$. The composition law of $\mathcal{G}$ can be depicted as
\begin{eqnarray}
M_{(a)}\circ M_{(b)}&\equiv&M_{(ab)}\label{MM}\\
N_{(a)}\circ N_{(b)}&\equiv&M_{(c)}\quad \mbox{with} \qquad c=ab\left(1+\xi^2\right)\quad\label{NN}\\
M_{(a)}\circ N_{(b)}&\equiv&N_{(ab)}\label{MN}
\end{eqnarray}

With this composition law, it is straightforward to show that they indeed form a group.
\begin{itemize}
\item[i)]Identity: ${\cal T}_1\circ {\cal T}_a={\cal T}_a\circ {\cal T}_1={\cal T}_a \quad \mbox{with} \quad {\cal T}_1=M_{(1)}$

\item[ii)] Inverse $M^{-1}_a=M_{a^{-1}}$ and $N^{-1}_a=N_{a^{-1}(1+\xi^2)^{-1}}$

\item[iii)] Closure is already given in (\ref{MM})-(\ref{MN})

\item[iv)] Associativity direct from (\ref{MM})-(\ref{MN})
\end{itemize}

The collection of all $M_{(a)}$'s forms a subgroup $\mathcal{H} \subset \mathcal{G}$. Actually, they are an invariant subgroup of $\mathcal{G}$ since
\[
N_{(a)}\circ M_{(b)} \circ N^{-1}_{(a)}=M_{(b)}\qquad .
\]

Thus, we can define an equivalence relation between the left coset defined in terms of the subgroup $\mathcal{H}$. Due to relations (\ref{MM})-(\ref{MN}) there is actually only two coset since
\[
\left[\mathcal{H}N_a\right]=\left\{N_c \, /\, c(x) \; \mbox{all analytical functions}\right\}
\]
In addition, the two cosets satisfy
\begin{eqnarray}
\left[\mathcal{H}\right] &\circ& \left[\mathcal{H}\right]=\left[\mathcal{H}\right]\\
\left[\mathcal{H}\right] &\circ& \left[\mathcal{H}N\right]=\left[\mathcal{H}N\right]\\
\left[\mathcal{H}N\right] &\circ& \left[\mathcal{H}N\right]=\left[\mathcal{H}\right]
\end{eqnarray}

Thus, the quotient group is $\mathcal{G}/\mathcal{H}=\mathds{Z}_2$

\subsection{The null Field $\kappa = 0$}

The algebraic null field, contrary to the general case, has an extra free function to define the disformal transformation. Relations (\ref{def_phi})-(\ref{phiphi}) show that a contravariant metric defined as
\[
\Delta^{\mu\nu}_{(a,b)}=a\ g^{\mu\nu}+b\phi^{\mu\nu} \quad ,
\]
has an inverse given by
\[
\Delta^{-1}_{(a,b)\mu\nu}=a\ g_{\mu\nu}-\frac{b}{a^2}\phi_{\mu\nu} \quad ,
\]
where the only condition over the functions $a$ and $b$ comes from (\ref{Deltak0}) that requires $a\neq 0$. The electromagnetic two-form in the $\Delta^{-1}_{(a,b)\mu\nu}$ geometry can be related with its counterpart in the $g_{\mu\nu}$ through
\[
F_{(a,b)}^{\mu \nu}=\Delta_{(a,b)}^{\mu\alpha}\Delta_{(a,b)}^{\nu\beta}F_{\alpha \beta}=a^2g^{\mu \alpha}g^{\nu\beta}F_{\alpha \beta}\quad.
\]
Therefore the relations (\ref{Fa})-(\ref{phia}) now modify to
\begin{eqnarray}
I_{1(a,b)}&=&a^2 I_1\label{Fab}\\
I_{2(a,b)}&=&a^2 I_2\\
\xi_{(a,b)}&=&\xi\\
\phi_{(a,b)\mu \nu}&\equiv&\Delta_{(a,b)}^{\alpha \beta}F_{\mu \alpha}F_{\nu \beta}=a\ \phi_{\mu \nu}\label{phiab}
\end{eqnarray}

In the same way, we define the transformation $\mathcal{T}$ acting on the metric $g_{\mu\nu}$ such that  
\begin{equation}\label{Transfk0}
\mathcal{T}_{(a,b)}[g_{\mu\nu}]\equiv \Delta^{-1}_{(a,b)\mu\nu} \quad .
\end{equation}

Applying a second transformation ${\cal T}_{(c,d)}$ on the metric $\Delta^{-1}_{(a,b) \mu\nu}$, we have
\begin{eqnarray}\label{Tk0}
{\cal T}_{(c,d)}\ [{\cal T}_{(a,b)}\ [g_{\mu\nu}]]&=&{\cal T}_{(c,d)}\ [\Delta^{-1}_{(a,b)\mu\nu}]\nonumber\\
&=&\Delta^{-1}_{(a.c\,,\ b.c+d.a^3) \mu\nu}\quad \Rightarrow \qquad \nonumber\\
\Rightarrow\quad {\cal T}_{(c,d)}\circ {\cal T}_{(a,b)}&=&{\cal T}_{(ac\ ,\ bc+a^3d)} \quad.
\end{eqnarray}

Therefore, the group properties are
\begin{itemize}
\item[i)]Identity: ${\cal T}_{(1,0)}$

\item[ii)] Inverse: ${\cal T}^{-1}_{(a,b)}={\cal T}_{(a^{-1},-ba^{-4})}$

\item[iii)] Closure: already given in (\ref{Tk0})

\item[iv)] Associativity: direct from the rule (\ref{Tk0})
\end{itemize}

Note that this is a non-abelian group, i.e.
\[
{\cal T}_{(c,d)}\circ {\cal T}_{(a,b)}\neq {\cal T}_{(a,b)}\circ {\cal T}_{(c,d)}\qquad.
\]

It is straightforward to show that the collection of all ${\cal T}_{(a,0)}$'s forms a subgroup. However, this is not an invariant subgroup since
\[
{\cal T}_{(c,d)}\circ {\cal T}_{(a,0)} \circ {\cal T}^{-1}_{(c,d)}={\cal T}_{\big(a\ ,\ dac^{-3}(a^2-1)\big)}\qquad .
\]

Notwithstanding, there is another subgroup $\mathcal{O}$ formed by the collection of all ${\cal T}_{(1,b)}$'s that are in fact an invariant subgroup. Indeed, we have
\[
{\cal T}_{(c,d)}\circ {\cal T}_{(1,b)} \circ {\cal T}^{-1}_{(c,d)}={\cal T}_{(1 ,bc^{-2})}\quad .
\]

Thus, we can establish an equivalence relation between the left coset defined in terms of this subgroup. The left cosets are of two types
\begin{eqnarray*}
\Big[\mathcal{O}_{1}\Big] &\equiv& \left\{\Delta^{-1}_{(1,b)} \ \Big \slash \ b \ \mbox{analytical}\right\}\\
\Big[\mathcal{O}_{c}\Big] &\equiv& \left\{\Delta^{-1}_{(c,d)} \ \Big \slash  c\neq 1\, \mbox{and} \ d \  \mbox{analytical}\right\}
\end{eqnarray*}

Contrary to the general case $\kappa \neq 0$, here there is an infinity of different left cosets labeled by the function $c$ above. Again, we can establish an equivalence relation between elements of the same left coset. These cosets inherit from the former group the following composition rule:
\begin{eqnarray}
\left[\mathcal{O}_1\right] &\circ& \left[\mathcal{O}_1\right]=\left[\mathcal{O}_1\right]\\
\left[\mathcal{O}_1\right] &\circ& \left[\mathcal{O}_a\right]=\left[\mathcal{O}_a\right]\\
\left[\mathcal{O}_a\right] &\circ& \left[\mathcal{O}_b\right]=\left[\mathcal{O}_{ab}\right]
\end{eqnarray}

Thus, the quotient group associated with disformal transformation in the the null case is infinite and characterized by all analytical 4-dimension real functions.\\

\section{Conclusion}

There is no question that symmetries play a fundamental role in modern physics. Noether's theorem, for instance, associate invariance of the action of a given physical system with conservation law's for tensor currents such as the energy-momentum tensor. In general, these symmetries are associated with variation of the action with respect to the spacetime coordinates and/or fields redefinitions.

In this paper, we have developed a different concept of symmetry that is closely related to the conformal symmetry. The point of departure is the definition of a dynamical system in an arbitrary spacetime $g_{\mu\nu}$ and the specification of the solutions of these equations of motion. Thereon, we have shown that the dynamics of the physical fields are invariant with respect to redefinitions of the metric tensors that maps different riemannian manifolds. More specifically, Maxwell's electrodynamics is invariant with respect to a large class of disformal metric transformations.

The disformal transformations together with the different manifolds that they generate has a group structure that is abelian for the algebraic general and non-abelian for the algebraic null case. There are several interesting issues related to the physical meaning of these disformally related manifolds and their mathematical structure that we shall address in future works.


\section{acknowledgements}
E. Goulart would like to thank FAPERJ for financial support.



\begin{thebibliography}{50}

\bibitem{ertov} F.T. Falciano, E. Goulart, \emph{A new symmetry of the relativistic wave equation}, Class.Quant.Grav. 29 (2012) 085011.
 
\bibitem{GNFT} E. Goulart, M. Novello, F.T. Falciano, J. D. Toniato, \emph{Hidden geometries in nonlinear theories: a novel aspect of analogue gravity}, Class. Quantum Grav. 28 (2011) 245008.

\bibitem{bek0} J. D. Bekenstein, in The Sixth Marcel Grossmann Meeting on General Relativity, ed.
H. Sato (World Publishing, Singapore, 1992).

\bibitem{bek} Jacob D. Bekenstein, The Relation between physical and gravitational geometry, Phys.Rev. D48 (1993) 3641-3647.


\bibitem{conf} Philippe Francesco, Pierre Mathieu and David Senechal, \emph{Conformal Field Theory} (Springer, 1997)

\bibitem{Kast} H.A. Kastrup, \emph{On the Advancements of Conformal Transformations and their Associated Symmetries in Geometry and Theoretical Physics, arXiv:0808.2730v1, 2008}

\bibitem{0} J. Magueijo, \emph{Bimetric varying speed of light theories and primordial fluctuations}, Phys.Rev. D79 (2009) 043525 

\bibitem{1} J. Magueijo, Rep. on Prog. in Phys. 66 (11), 2025, 2003. 

\bibitem{2} J. W. Moffat, Int. J. Mod. Phys. D 2, 351-366 (1993). 

\bibitem{er} E. Goulart, Santiago Esteban Perez Bergliaffa, \emph{Effective metric in nonlinear scalar field theories}, Phys.Rev. D84 (2011) 105027. 


\bibitem{san} R.H. Sanders, \emph{Solar system constraints on multi-field theories of modified dynamics}, Mon.Not.Roy.Astron.Soc. 370 (2006) 1519-1528. 


\bibitem{Tev} Constantinos Skordis, \emph{The Tensor-Vector-Scalar theory and its cosmology},  Class.Quant.Grav. 26 (2009) 143001.

\bibitem{Mass}Valentina Baccetti, Prado Martin-Moruno, Matt Visser, \emph{Massive gravity from bimetric gravity}, Class.Quant.Grav. 30 (2013) 015004. 

\bibitem{dbi} Miguel Zumalacarregui, Tomi S. Koivisto, David F. Mota, \emph{DBI Galileons in the Einstein Frame: Local Gravity and Cosmology}, IFT-UAM-CSIC-12-C3, 2012. Claudia de Rham  \emph{Galileons in the Sky}, Comptes Rendus Physique 13 (2012) 666-681.

\bibitem{disform} Nemanja Kaloper, \emph{Disformal inflation}, Phys.Lett. B583 (2004) 1-13. 

\bibitem{new} M. Zumalacarregui, T.S. Koivisto, D.F. Motac and P. Ruiz-Lapuentea, \emph{Disformal scalar fields and the dark sector of the universe}, JCAP 05, (2010), 038.

\bibitem{causal} Erico Goulart de Oliveira Costa, Santiago Esteban Perez Bergliaffa, \emph{A Classification of the effective metric in nonlinear electrodynamics}, Class.Quant.Grav. 26 (2009) 135015  

\bibitem{synge} J. L. Synge, \emph{Relativity: The Special Theory}, (North-Holland Publishing Company, 1958).

\bibitem{Had} Y. Choquet-Bruhat, C. de Witt-Morette, and M. Dillard-Bleick, \emph{Analysis, Manifolds and Physics} (North-Holland, New York, 1977) p. 455.

\end{thebibliography}
\end{document}